\documentclass{article}

\usepackage{arxiv}

\usepackage[utf8]{inputenc} 
\usepackage[T1]{fontenc}    
\usepackage{hyperref}       
\usepackage{url}            
\usepackage{booktabs}       
\usepackage{amsfonts}       
\usepackage{nicefrac}       
\usepackage{microtype}      
\usepackage{lipsum}
\usepackage{graphicx}

\usepackage[natbibapa]{apacite}
\title{How do climate change skeptics engage with \\ opposing views? \\
    \vspace*{0.5cm}
  \large Understanding mechanisms of social identity and cognitive dissonance in an online forum}

\author{
  Lisa ~Oswald\thanks{OrcidID: https://orcid.org/0000-0002-8418-282X, Twitter: @LisaFOswaldo} \\
  Hertie School \\ 
  Friedrichstraße 180 \\ 
  10117 Berlin, Germany \\
  \texttt{l.oswald@phd.hertie-school.org} \\

   \And
 Jonathan ~Bright \\
  Oxford Internet Institute \\
  University of Oxford \\
  1 St Giles', Oxford OX1 3JS, UK \\
  \texttt{jonathan.bright@oii.ox.ac.uk} \\
}

\begin{document}

\maketitle

\begin{abstract}
Does engagement with opposing views help break down ideological `echo chambers'; or does it backfire and reinforce them? This question remains critical as academics, policymakers and activists grapple with the question of how to regulate political discussion on social media. In this study, we contribute to the debate by examining the impact of opposing views within a major climate change skeptic online community on ‘Reddit’. A large sample of posts (N = 3000) was manually coded as either dissonant or consonant which allowed the automated classification of the full dataset of more than 50,000 posts, with codes inferred from linked websites. We find that ideologically dissonant submissions act as a stimulant to activity in the community: they received more attention (comments) than consonant submissions, even though they received lower scores through up-voting and down-voting. Users who engaged with dissonant submissions were also more likely to return to the forum. Consistent with identity theory, confrontation with opposing views triggered activity in the forum, particularly among users that are highly engaged with the community. In light of the findings, theory of social identity and echo chambers is discussed and enhanced. 
\end{abstract}


\section*{Introduction}
Scientists worldwide consider climate change as one of the greatest challenges of our time \citep{ipcc_ar5_2015}. Its dynamics, consequences and roots in human activity attract enormous scientific consensus \citep{cook_consensus_2016}. However, despite this consensus public skepticism about climate change remains significant in many countries \citep{poberezhskaya2018blogging}. For example, only 29\% of U.S. conservatives attribute climate change to human activity \citep{funk2015chapter}, whilst the European Social Survey shows that climate change skepticism is expressed by 3 to 10\% of respondents in Western Europe. In Eastern European countries such as the Czech Republic, Lithuania and Estonia, as well as in Israel and in Russia, such skepticism is present in the answers of 10-15\% of respondents \citep{poortinga2019climate}. Findings from a representative UK study \citep{poortinga_uncertain_2011} and the Eurobarometer \citep{mccright_political_2016} reinforce these results. Such skepticism inevitably has the potential to impede both action on climate change in particular \citep{runciman_how_2017} and also trust in science more generally. \citep{almassi_climate_2012,beck_between_2012,hmielowski_attack_2014}.

While climate change skepticism has a long history, it currently often manifests itself in debates and discussions on social media (see, for example, studies by \citealt{pearce_social_2019}, \citealt{williams_network_2015}, \citealt{tyagi2020polarizing}, \citealt{samantray2019credibility} and \citealt{treen2020online}). Climate change skeptics have the opportunity to gather and organise in online forums, emphasising uncertainties about scientific findings, spreading contrary opinions and weakening the support for political action on mitigation and adaptation strategies. This largely informal and non-institutional communication often excludes robust scientific information \citep{schafer_online_2012}. Instead, those skeptical about climate change can establish a supportive, networked space linked to other climate change skeptical sites; one that distances readers from original sources of scientific information. Within this network, users engage in a variety of rhetorical strategies to echo skeptical opinions and to discredit opposing views \citep{bloomfield2019circulation}. In such spaces, climate skepticism acts as a local majority: and such `perceived consensus' has strong potential to affect opinion formation  \citep{lewandowsky2019science, marks1987ten}.

While many researchers in the field of such climate change `echo chambers' \citep{sunstein_law_2002} identify patterns of polarization but refrain from giving clear recommendations on how to prevent or mitigate them \citep{bruggemann2020mutual,elgesem2015structure,kaiser2017alliance}, current work has started to propose interventions that promote engagement with cross cutting views and opinions as a way of undermining this type of ideological enclave \citep{edwards2013participants,nordbrandt2020cross,van2020echo,walter_echo_2018,williams_network_2015}. However, more recent research in other contexts has suggested that such exposure may backfire \citep{bail_exposure_2018,paluck_is_2010}. Furthermore, experimental evidence shows that skepticism is a complex issue which is hard to address or `debunk' \citep{cook_quantifying_2013}. Communication attempts that focus on the scientific consensus \citep{myers2015simple,van2014communicate}, or that use targeted messages that resonate with audience values such as religion or a free market ideology \citep{campbell2014solution,dixon2017improving} offer some potential. However, \citet{dixon2017improving} did not find significant effects for scientific consensus messaging, and overall, backfire or boomerang effects \citep{lodge2013rationalizing} can increase climate change skepticism within these attempts \citep{cook_consensus_2016,hart2012boomerang,zhou2016boomerangs}. Hence, overall the effect of cross cutting and opposing information on climate change skeptic views remains highly unclear. 

In this study, we build on this previous literature with an observational study of how climate change skeptics react engage with opposing viewpoints. Our work is based on theories of social psychology and social cognition, which offer diverging expectations about the consequences of dissonant material when it enters an echo chamber. The empirical part of our work looks at r/climateskeptics, a part of Reddit that is dedicated to the critique of the concept of climate change. Our results show the nuanced ways in which individuals in an echo chamber are able to manage and cope with the dissonance they encounter towards their views. 

\section*{Theorising responses to opposition in climate change skeptic communities}

We ground our study in social identity theory. This theory deals with the portion of an individual’s self-concept that is derived from perceived membership of a social group \citep{tajfel_integrative_1979}. It argues that inter-group contact, in contrast to the literature on within group polarisation \citep{myers1976group,cialdini_social_2004,yardi2010dynamic,jones2000potential}, has the potential to intensify the need to maintain one’s identity by identifying with the ingroup and arguing against the outgroup \citep{hogg2007uncertainty,tajfel_integrative_1979,turner_social_1998}. The way individuals categorise the world into mental schemes is the core principle of social cognition \citep{fiske_social_1991} that can lead to ingroup preference, stereotypes and logic errors \citep{aronson_social_2014}. 

Following the suggestion of \citet{jost_ideological_2014}, who actively call for the incorporation of theories of motivated social cognition into studies of information-exposure, in this paper we test two lines of thinking about the nature of responses to opposing viewpoints within a climate change skeptic community: mechanisms of selective exposure and mechanisms of social identity and motivated social cognition. Such theories, it is worth highlighting, have been widely applied in previous studies of climate change skeptic communities \citep{douglas_psychology_2017,hakkinen_ideology_2014,corner2012uncertainty}, which have often connected such skepticism to issues of identity rather than simple lack of understanding or information \citep{hogg_uncertainty_2014,hogg_uncertaintyidentity_2013}. Such works build on the observation that there are often gaps between environmental attitudes and behaviour \citep{bamberg_twenty_2007,kollmuss_mind_2002}, and gaps between information and attitude \citep{arcury_environmental_1990,bradley_relationship_1999} in the context of climate change.

The empirical component of our paper is based on the social news aggregation, content rating and discussion platform `Reddit'. On Reddit, users are registered under a pseudonym. They are able to create `submissions' in a large number of `subreddits', which are parts of the site dedicated to the discussion of specific topics. These submissions are often links to other websites (though can also be text, images, or videos). Other users can then read, up-vote, down-vote, and comment on the submission. Reddit is a highly frequented website and most content is publicly available which makes it an interesting resource of organic content for social scientists \citep{amaya_new_2019}. It is also a good case for our study because it contains a large climate change skeptic subreddit: r/climateskeptics. This subreddit is the focus of our study. It represents, arguably, an echo chamber \citep{sunstein_law_2002}: a place where a radical, counter-cultural worldview which is a minority view in society at large can nevertheless thrive and be in a local majority \citep{bright_echo_2020}. In what follows below, we elaborate theoretical expectations for the types of reactions dissonant viewpoints might provoke in the context of such a space. 

We will begin by looking at the type of direct responses that dissonant submissions might provoke. In line with theories of social cognition, it is likely that members of a community rate opinions and information provided by  members of their ingroup higher than content coming from an outgroup. This is due to mechanisms of ingroup favoritism and negative outgroup stereotypes \citep{zebrowitz2007contribution}, as well as mechanisms of motivated social cognition \citep{jost2012political}. This is also supported by the theory of cognitive dissonance \citep{festinger_theory_1957,greenwald_twenty_1978} which states that information or media messages that challenge people’s beliefs typically create a feeling of dissonance, which is unpleasant and something most people avoid \citep{hart_feeling_2009}. Cognitive dissonance can be reduced by (amongst other methods) reducing the importance of dissonant cognitions, or increasing the importance of consonant cognitions \citep{harmon-jones_introduction_2019}.

In our particular case, the critical method for indicating the quality of a submission in Reddit is voting: users can `upvote' submissions they agree with or think are important, and `downvote' others. Upvotes and downvotes are combined together to form a score\footnote{The sum of positive up-votes and negative down-votes results in the `score' of a submission.} for each individual submission, which is a critical part of how Reddit organises the presentation of content to users. Hence, building on the theories outlined above, we expect that the importance of a dissonant submission could be lowered by undermining its legitimacy through `downvoting' it. Conversely, such opposing submissions can be further undermined by `upvoting' consonant ones. We hence develop our first hypothesis: 

\paragraph{Hypothesis 1:} Submissions featuring opposing views and dissonant information will receive lower scores than consonant ones. 
\hfill \break

The theory of motivated reasoning \citep{jost_exceptions_2003} assumes that attitude strength is an important moderator for cognitive biases. \citet{karlsen_echo_2017} found that to some extent, people seek supportive messages, but that it is more difficult to show that people avoid contradictory information \citep{brundidge_encountering_2010}. The theory further assumes that prior attitudes bias how people process arguments, and that this bias is reinforced not only through selective exposure, but also through selective judgement processes \citep{lebo_aggregated_2007,taber_motivated_2009,taber_motivated_2006}. The first mechanism, the confirmation bias or attitude congruency bias \citep{taber_motivated_2009}, assumes that people tend to evaluate arguments that support their views as strong and compelling, which would be in line with the argument that echo chambers lead to polarised opinions. The second mechanism is disconfirmation bias, according to which people use time and cognitive resources to degrade and counter argue opposing views \citep{taber_motivated_2009}. Therefore, when presented with contrasting arguments in online debates, these arguments may lead to even stronger beliefs in initial opinions. For example, \citet{karlsen_echo_2017} found that people engaged actively with dissonant content from the outgroup and that their opinions were reinforced by contradiction as well as confirmation.

In our particular context, apart from simply voting, contradiction can also be signalled by commenting on a submission. Of course, comments can be positive or negative, however we expect that dissonant posts provide a stronger motivation to respond than consonant ones, based on the reasoning above (it is worth noting that our twin hypotheses of lower score and higher comments goes against the general logic of Reddit, in which score and comments are typically positively correlated - see \cite{singer2016evidence}). Therefore, we develop our second hypothesis:

\paragraph{Hypothesis 2:} Submissions featuring opposing views and dissonant information will trigger more activity (comments) than climate change skeptic ones. 
\hfill \break

The temporal Need-Threat Model, originally developed in the context of social ostracism \citep{williams_ostracism_2009}, assumes that when fundamental needs (such as the needs for belonging, self-esteem, meaning, and control) are threatened, people initially react in a quick and reflexive way that includes strategies of need fortification. Needs for belonging and self-esteem can be fortified through attempts to become more socially attractive. This can be achieved by being attentive to social cues, as well as overall compliance and conformity. A threatened need for control and recognition can be rebalanced through provocation, gaining attention, lashing out, and retaliation. Considering these mechanisms in the context of online social platforms, engagement with information that strongly questions the worldview and identity of a user could potentially threaten needs for meaning, self-esteem and control. In addition to the demonstration of conformity by posting or up-voting climate change skeptic content, a provoking and aggressive response could also work as reaction mechanism to fortify threatened needs. Given this, we state our third hypothesis:

\paragraph{Hypothesis 3:} Submissions featuring opposing views and dissonant information are more likely to attract comments expressing negative sentiment.
\hfill \break

However, if these attempts to undermine or contradict dissonant information are unsuccessful people can reach a resignation stage \citep{williams_ostracism_2009}. With increasing frustration through ongoing inability to fortify needs of control and self-esteem, social withdrawal can occur. Another possible coping mechanism can be engagement with other, more pleasant activities. This form of distraction is conceptualized as an accommodative, secondary control coping strategy \citep{allen2010self}. 
In the case of online communities, we might imagine that, following engagement with a dissonant submission, people would possibly suspend their engagement with a forum and move on to something else, which would again align with the tendency to avoid information that induces cognitive dissonance \citep{case2005avoiding}. In contrast to hypotheses 1 to 3, this effect is in line with the mechanisms of selective exposure in online echo chambers, where people are predominantly seek exposure to attitude, control and self-esteem reinforcing information.

\paragraph{Hypothesis 4:} Engagement with opposing views and dissonant information makes users more likely to leave the forum.  
\hfill \break

It is also worth considering the type of user that responds to dissonant content. It is by now well known that the distribution of activity in online forums is skewed: a highly active minority typically accounts for a significant percentage of all the activity taking place \citep{bright_power_2019}. In the work of \citet{barbera_tweeting_2015}, these levels of contribution activity was also related to political extremism, suggesting that these highly active individuals also hold stronger views. The majority of content was created by a minority of extreme conservatives and extreme liberals; and the relationship between extremism and the formation of echo chambers has also been documented in other work \citep{bright_explaining_2018}.  

According to Social Identity Theory \citep{tajfel_integrative_1979}, the structure of an individual’s social identity depends on the relevance of perceived membership of a social group or community. Often, an individual is a member of various social groups, such as a family, a work team, sports teams, etc. In the digital sphere, this spectrum expands to online communities. Furthermore, the more strongly people engage with a community, the more relevant is this membership for their social identity. In case identification with one particular social group dominates the social identity concept of an individual, a threat to the identification with this group can pose substantial stress on the individual \citep{haslam_maintaining_2008,praharso_stressful_2017}. In the case of power users, who spend a considerable amount of time engaging with an online community, their social identity may be more strongly affected by external threats to or dissonances within the online community, compared to users whose social identity depends less on their identification with the online community due to a lower level of engagement. In line with the Need-Threat Model, this identity threat, or in other words, the threatened need to belong, can be dealt with in various forms, such as the active confrontation or counter argument against users who threaten the consistency of community beliefs when injecting opposing views and dissonant information into the community.

\paragraph{Hypothesis 5:} More senior users, those with higher levels of past community engagement, are more likely to engage in discussions in reaction to submissions featuring opposition to climate change skepticism, compared to users with a lower level of past engagement with the community.

\section*{Methods} 

\subsection*{Data Collection}

Data for the study was collected from the r/climateskeptics subreddit\footnote{See: \url{https://www.reddit.com/r/climateskeptics/}}. This subreddit was created in July 2008 and encourages skepticism and debate about the concept of climate change. It contains more than 50,000 submissions on the topic, the majority of which (92\%) are simply links to websites, blogs, articles or videos. These submissions have, at the time of writing, attracted more than 350,000 comments, made by over 13,000 users. The subreddit is open to everyone, but the content is moderated by four users with the right to delete submissions and comments and to `ban' users from the subreddit. No clear guidance is set out in terms of how submissions might be selected for deletion. The only rule explicitly mentioned in the description of the subreddit is that a `disparagement of the subreddit as a whole is a bannable offense'. However, criticism of climate change skeptics (dissonant viewpoints, in our terms) is a consistent feature of the forum. 

The data for the paper was collected from two different APIs. All submissions were collected through pushshift.io \citep{baumgartner_pushshift_2018}, which is known to provide a near complete repository of Reddit submission data \citep{baumgartner2020pushshift}. Once submissions had been collected, further metadata and comments were collected through the PRAW Reddit API  \citep{reddit_inc_api}. This next step revealed that 8.3\% of submissions (4,213) available through Pushshift have since been deleted. However, we are still able to include these submissions in our analysis as pushshift retains information about whether they contained a link or not, and if so what it was (which, as we will describe below, is the critical part of our coding methodology). We include a dummy variable indicating whether the submission was deleted or not as a control variable in our models (clearly, submissions which were swiftly deleted would be less likely to attract comments or gain a high score). We manually reviewed a random selection of 40 submissions that had not been deleted, and 129 associated comments, and found no missing data problems. 

The data was stored securely and usernames were pseudonymised. All results are reported only in aggregate to preserve the anonymity of those who had made use of the forum. The study received approval from the Oxford Internet Institute's internal research ethics committee (reference: SSH OII CIA 20 015). 

\subsection*{Measures}

The most important measure in our study concerns whether a submission to r/climateskeptics should be labelled `consonant' or `dissonant'. A consonant submission, in this context, would be one that supports the majority opinion on the forum by putting forward views that are skeptical  of climate change. A dissonant submission, by contrast, would be one that supports the scientific consensus on climate change \citep{cook_consensus_2016} and attacks or undermines climate change skeptics. This dichotomy is of course a simplification of the range of views someone can have on climate change: for example many of the people in r/climateskeptics were simply skeptical about some of the aspects of climate science such as the anthropogenic cause or the negative consequences for humans rather than denying outright the warming trend in the earth's climate (see \citet{leiserowitz_climate_2020} for an overview of different styles of reasoning on climate change). Nevertheless, this simplification captures the essence of what we want to study. 

We approached this labelling task in two ways. First, we took a randomly drawn sample of 3,000 submissions, and one author of the study manually coded each one as either `consonant' or `dissonant'. Such a manual approach seemed important because it allows for the interpretation of ironic or sarcastic content, image material or the disapproving reference of climate activist content, which would be mislabelled using an automated or purely semantic approach \citep{metag_content_2016,grimmer_text_2013,riffe_analyzing_2019}. A second author double coded a randomly drawn sample of 100 of these submissions, to calculate an estimate for the inter-rater reliability of the coding. Overall there was an 86\% agreement between coders, with a Krippendorff's alpha of 0.62, indicating a substantial accordance \citep{mchugh_interrater_2012}.  

Second, we leveraged the manual coding to assign an automatic dissonance code to every submission in the dataset. We did this by exploiting the fact that 92\% of the submissions in our dataset were simply links to external sites, as described above. We used the data from our manually coded sample to infer a code for these external sites, by taking an average of codes from submissions that contained this link in our manually coded sample. These inferred codes can then be applied to other submissions in the dataset. For example, the Daily Caller appeared 46 times in our manually coded data. Two of these submissions were coded as dissonant, and 44 of them were coded as consonant. Therefore, in our automatically coded data, all submissions containing the Daily Caller received an automatic dissonance code of \(\frac{2}{46} = 0.04\). One might question how some sites can be used to advance both consonant and dissonant positions. However, one thing we remarked on in the coding was how easy it was to use a link from an outlet to support whatever position the author of the submission wanted. For example, the Guardian is a site that clearly supports the idea that climate change is a substantial threat \citep{carrington_2019}. It featured 40 times in our data, however the outlet received an average dissonance code of 0.175 in our data, which means it was largely used to support climate change skeptical positions. 

Using these methods, we were able to infer an automatic dissonance code for 83\% of the submissions in our dataset. The remaining 17\% of submissions could not be assigned a code in this way, either because they did not contain a link, or because they contained a link which was not contained in our manually coded data. These submissions were given dissonance codes based on the average dissonance code of other submissions the author had made, on the basis that author were likely to be relatively consistent in their beliefs. If even that was not possible (if the author themselves had no other coded submissions), then the submission was given a code based on the average dissonance code of all other authors (see Appendix \ref{app:imputation} for more details). As part of our diagnostic checks, we checked the sensitivity of our results to this method of imputation, as described in Appendix \ref{app:diagnostics}. We found no evidence that the imputation substantially affected the results. 

We also coded the sentiment present in the comments in our dataset. We made use of a widely used approach, which  determines the sentiment of a text section as the sum of the sentiment content of the individual words \citep{silge2016tidytext}. Out of various general-purpose dictionaries that encode sentiment for single words, we chose the NRC sentiment dictionary \citep{mohammad2013crowdsourcing}. This dictionary, constructed using crowdsourcing, categorises words as positive or negative. The NRC sentiment dictionary was validated on online communication data such as restaurant or movie reviews \citep{kiritchenko2014nrc} and conversations on Twitter \citep{mohammad2013nrc} which is relatively close to the type of content we face in our project. Once we had applied the NRC dictionary to the text of the comments, the resulting sentiment for each comment was defined simply as the number of positive words minus the number of negative words. Of course, it is worth noting that sarcasm or negated text can be a challenge for the validity of automated sentiment analysis \citep{silge2016tidytext}.

In addition to these coding exercises we also collected a number of other variables which were relevant for the study. We recorded the `score' of each submission. This score is simply the number of up-votes minus the down-votes a submission receives. With this metric, absolute votes are concealed\footnote{For example, a score of 2 could mean that a submission received 3 up-votes and 1 down-vote, or that it received 50 up-votes and 48 down-votes.} to mitigate the influence of spam bots \citep{reddit_inc_score}. We recorded the number of comments a submission received, and the overall number of contributions (submissions and comments) by each user in the dataset. Finally, we recorded the timestamp of the submission and the number of comments in reaction to the submission. All of these pieces of data were collected from the PRAW API.

\begin{table}
\caption{Descriptive statistics} 
\label{tab:desc_stats} 
\centering
\resizebox{0.6\textwidth}{!}{\begin{tabular}[t]{lcc}
\toprule
& Manual labelling & Automatic labelling \\
\midrule
Dissonance (mean) & 0.094 & 0.095 \\
Dissonance (sd) & 0.292 & 0.151 \\
Score (mean) & 14.21 & 13.73  \\
Score (sd) & 31.79 & 30.31 \\
Num. comments (mean) & 6.88 & 7.02 \\
Num. comments (sd) & 14.65 &  14.74 \\
Comment sentiment (mean) & 0.49 & 0.54 \\
Comment sentiment (sd) & 1.47 & 1.51 \\
\% Text only & 92\% & 92\% \\
\% Deleted & 92\% & 92\% \\
Num.Obs. & 3,000 & 50,502 \\
\bottomrule
\multicolumn{3}{p{0.6\textwidth}}{\textit{Note.}`Dissonance' in relation to climate change skepticism; `score' as sum of upvotes and downvotes; `text only' representing submissions without links to external websites; `deleted' representing submissions which were subsequently removed.} \\
\end{tabular}}
\end{table}

Descriptive statistics for the measures are shown in Table \ref{tab:desc_stats}. The average dissonance code is low, as we would expect in a forum which is largely dedicated to the idea of climate change skepticism. The dissonance codes are also quite consistent between the manual and automatic labelling. The metrics for score, the number of comments and sentiment all show skewed distributions, which suggests they should be log transformed before analysis. 

\section*{Analysis}

\begin{table}
\caption{Immediate reaction to dissonant content} 
\label{tab:m1-3} 
\centering
\resizebox{0.7\textwidth}{!}{\begin{tabular}[t]{lcccccc}
\toprule
 & \multicolumn{2}{c}{Score} & \multicolumn{2}{c}{Comments} & \multicolumn{2}{c}{Sentiment}\\
 & M1.1 & M1.2 & M2.1 & M2.2 & M3.1 & M3.2\\
  & (Manual) & (Auto) & (Manual) & (Auto) & (Manual) & (Auto)\\
\midrule
Dissonance & 0.32*** & 0.29*** & 2.00*** & 1.77*** & 1.00 & 1.00\\
Text Only & 0.61*** & 0.70*** & 1.80*** & 1.88*** & 1.01 & 1.01***\\
Deleted & 0.55*** & 0.54*** & 0.62*** & 0.62*** & 1.00 & 1.00\\
\midrule
Num.Obs. & 3,000 & 50,502 & 3,000 & 50,502 & 2,140 & 35,730\\
Adj. $R^2$ (full)  & 0.277 & 0.203 & 0.139 & 0.119 & -0.001 & 0.004\\
Adj. $R^2$ (proj.) & 0.143 & 0.080 & 0.056 & 0.044 & -0.020 & -0.000 \\
\bottomrule
\multicolumn{7}{l}{\textsuperscript{} * p $<$ 0.05, ** p $<$ 0.01, *** p $<$ 0.001}\\
\multicolumn{7}{l}{All coefficients are exponentiated}\\
\end{tabular}}
\end{table}

In Table \ref{tab:m1-3} we present our first set of results, examining the reactions generated by dissonant content within the r/skeptics subreddit. Each model in Table \ref{tab:m1-3} is a fixed effects linear regression, and each observation in the model is a different submission made to the forum. We include fixed effects on the year, day of week and hour of posting to control for temporal effects in the data, as both submission and comment levels vary strongly over time. We also include whether the post is a `text only' post (and contains no link within it) and whether the post has been deleted as control variables. We use HC1 robust standard errors for all models \citep{long2000using}. Standard diagnostic checks were applied which provided no reason to doubt the results. A full record of diagnostics can be found in Appendix \ref{app:diagnostics}. Each outcome variable is log transformed. All coefficients are exponentiated, and hence can be interpreted as percentage changes. 

We will begin with hypothesis 1, which suggested that dissonant submissions ought to receive a lower score than consonant ones. Models 1.1 and 1.2 test this proposition using our manually and automatically coded data. In both our manual and automatic models, dissonant submissions received around 70\% lower scores than submissions confirming climate change skepticism, hence providing strong supporting evidence for hypothesis 1.

Hypothesis 2 suggested that dissonant submissions ought to receive more comments. Again we find support for this idea in both our manual and automatic models: submissions undermining climate change skepticism received 100\% more comments than those which do not in our manually coded model, and 77\% more in the automatically coded model. Hence hypothesis 2 is also supported. 

Hypothesis 3 suggested that comments on dissonant submissions would have a more negative sentiment than comments on submissions supporting climate change skepticism. Here we find no evidence of any relationship in either our manual or automatic models, and an extremely low R\textsuperscript{2} value for both of them. So hypothesis 3 is not supported. It is worth noting that M3.1 and M3.2 can only be estimated for submissions that attracted at least one comment. Hence, the N is lower than in models M1.1 - M2.2.

One point worth noting across all models is the the consistency of results between our manual and automatic approaches. This suggests that our automatic approach offers a good proxy for the more reliable but smaller scale manual coding.  

Our next question concerned what happens after people have engaged with dissonant material, which we define here as having commented on a dissonant submission. We hypothesised (H4) that those having their beliefs challenged in this way would be less likely to return to the forum than those who had engaged with attitude confirming information. We address this question using a PWP-Gap Time model, a type of event history analysis \citep{amorim2015modelling}. This model (M4.1) is presented in Table \ref{tab:m4}. Each observation in the model is a comment to the subreddit made by one of the authors in the dataset on one of the submissions. By looking at the characteristics of the submission which attracted the comment, and the time elapsed until the next comment, we can analyse how engagement with either consonant or dissonant submissions relates to the probability of returning to the forum and commenting again. 

We only present a version of this model using automatically coded data, as it presupposes complete comment histories for each author, which of course cannot be achieved with the manual data. The model clusters standard errors at the author level, takes into account the number of previous comments made by each author, and also stratifies according to the hour, weekday and year of contribution. The number of comments made on the submission, and the average sentiment of the comments, are included as control variables. Diagnostic checks for this model can be found in Appendix A2: they also provide no reason to doubt the results.

\begin{figure}[ht]
\includegraphics[width=0.9\linewidth]{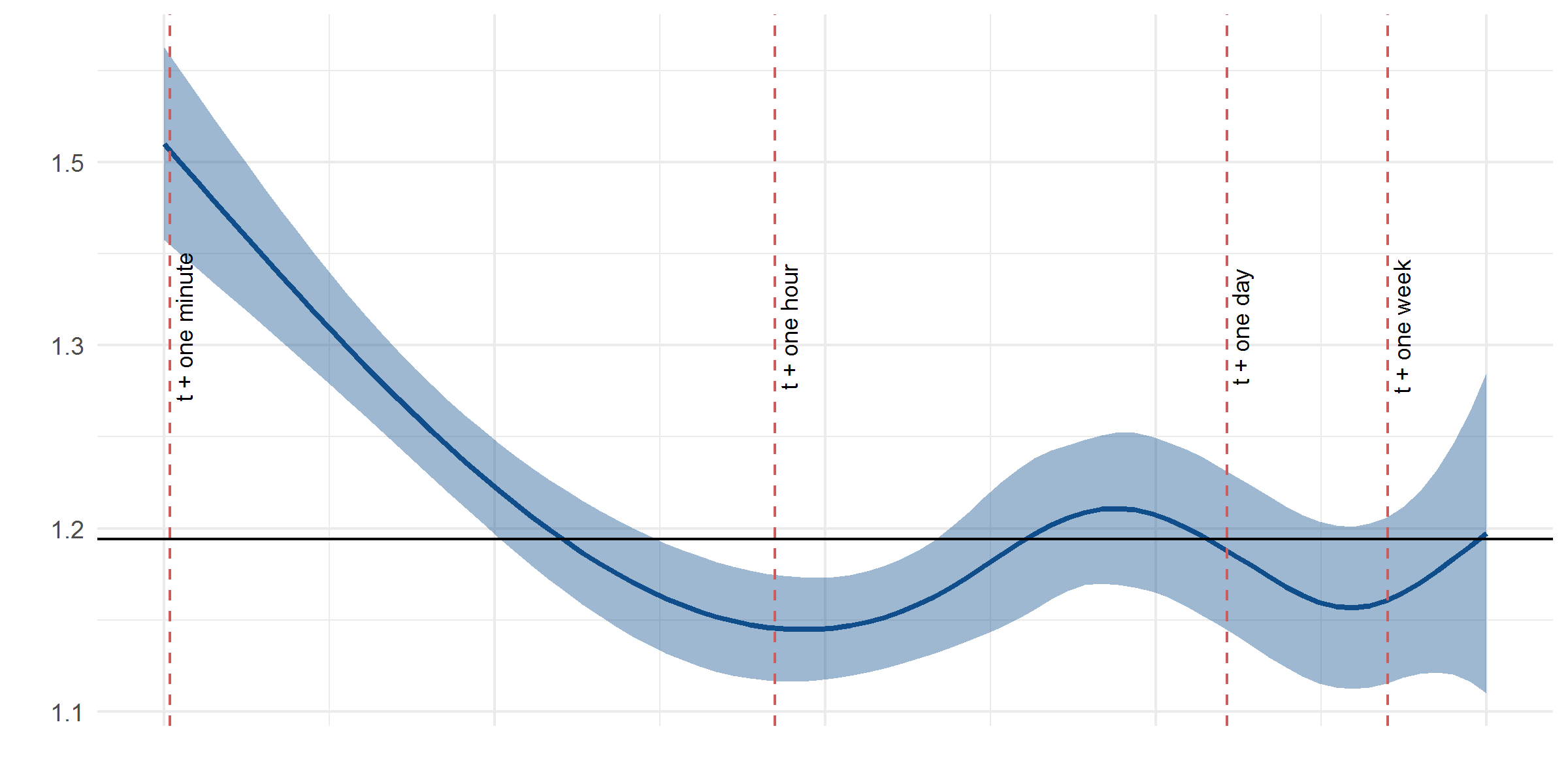}
\caption{Variation in effect of dissonant engagement on relative chance of commenting again over time. Horizontal line indicates average effect reported in M4.1} 
\label{prop_haz}
\end{figure}

The results of the model show that our hypothesis is reversed: those engaging with dissonant material were around 21\% more likely to post again compared to those who did not. Having commented on submissions which attracted more comments or which had more positive sentiment also increased the likelihood of coming back. A proportional hazards test for the model shows that the effect varies over time. We show this variation in Figure \ref{prop_haz}. The horizontal line in the figure indicates the average effect reported in M4.1. We can see that the effect of having engaged with a dissonant post stimulates a much stronger likelihood of coming back in the first hour or so (starting from around 50\% more in the first minute). This then declines progressively over time. It is also worth commenting on the very low Cox-Snell R\textsuperscript{2} observed: clearly the model has low explanatory power for understanding the full reasons why some people are more likely to return to a forum than others.

\begin{table}
\caption{Seniority of those engaging with dissonant content, and their next destination} 
\label{tab:m4} 
\centering
\begin{tabular}[t]{llll}
\toprule
  & Prob. post again & \multicolumn{2}{c}{Seniority}\\
  & \multicolumn{1}{c}{M4.1} & \multicolumn{1}{c}{M5.1} & \multicolumn{1}{c}{M5.2} \\
  & \multicolumn{1}{c}{(Auto)} & \multicolumn{1}{c}{(Manual)} & \multicolumn{1}{c}{(Auto)} \\
\midrule
Dissonance  & 1.21*** & 1.38*** & 1.43***\\
Text Only   & 1.18*** & 1.47*** & 1.53*** \\
Deleted        & 0.97  & 0.91 & 0.96*   \\
Avg. sentiment comments & 1.03***\\
Num. comments (log)  & 1.24***\\
\midrule
Num.Obs.     & 332,014 & 2,101 & 35,040      \\
Adj. $R^2$(full) & & 0.097 & 0.095         \\
Adj. $R^2$(proj.) & & 0.021 & 0.033  \\
Cox-Snell $R^2$    & 0.006     \\
\bottomrule
\multicolumn{4}{l}{\textsuperscript{} * p $<$ 0.05, ** p $<$ 0.01, *** p $<$ 0.001}\\
\multicolumn{4}{l}{All coefficients are exponentiated}\\
\end{tabular}
\end{table}

In our final set of results we investigate the type of people who engage with dissonant submissions in the forum. In particular, we hypothesised (H5) that those engaging with such submissions would be more likely to be senior members of the forum. Models M5.1 and M5.2 in Table \ref{tab:m4} test this suggestion. Both are linear fixed effects models, and are specified in a similar way to models 1-3: each observation is a submission. The dependent variable is the mean number of previous contributions (both comments and submissions) that people commenting on this submission had made to the subreddit at the time of commenting. We should note that we can only estimate this for submissions which attracted comments from users who had not deleted their Reddit account at the time of writing. Hence, the N is slightly lower than it is in M3.1 and M3.2.

In the models, we find strong support for H5: in both our manual and automatic models, those commenting on dissonant submissions have made around 40\% more previous contributions than those commenting on consonant submissions. In other words, more dissonant posts attract more senior figures to respond.

\section*{Discussion}

This study analysed the consequences of opposing views and dissonant information within a major climate change skeptic online community. A number of findings are worth emphasising. First, we showed that dissonant submissions to the forum attracted a lower score but a higher amount of comments. This finding, while expected from the literature we elaborate above, is an interesting reversal of general trends on Reddit where score and comments tend to be correlated \citep{singer2016evidence}. One thing this shows is how engagement in online discussion forums provides sophisticated tools for managing dissonance: individuals in the forum can at once undermine the credibility of the author (by down-voting it) but also engage strongly with it (by commenting). In a way, dissonance can be micro-managed in such settings, which (we speculate) may make minimising it easier. 

Furthermore, there is a tendency for more senior users to be especially engaged within the discussions in reaction to submissions that contain opposing views and dissonant information. This tendency hints towards the hypothesised effect of identity defence. Users highly engaged with r/climateskeptics might derive a greater part of their social identity from the identification with this social group. Consequentially, their experienced identity threat when confronted with opposing views leads to strategies of identity and community defence, in the form of argument countering engagement in the comments below an opposing post. Given that identity strength is an important moderator for motivated reasoning \citep{jost_exceptions_2003}, our results align with the theory: those highly engaged with the forum are more likely to engage in discussions triggered by opposing views. This is also in line with the findings suggesting that contribution activity is related to political extremism \citet{barbera_tweeting_2015}. However, the study of \citet{barbera_tweeting_2015} did not differentiate between activity levels after engagement with dissonant and congruent information.

Against our expectation of the long term effects of need threats, users who engaged with opposing views were more likely to return to the forum than those engaging with attitude confirming skeptic content. However, given that the resignation stage proposed by the Need-Threat model might not be reached, this finding actually aligns with the expectations from social identity theory. Identity threats produced by confrontation with opposing views may in fact increase the tendency to defend one's identity and to reduce feelings of dissonance by returning to the forum and posting identity reinforcing content into the subreddit. The fact that dissonant viewpoints are fundamental for stimulating activity in the forum and encouraging people to return may prompt a rethink of the overall idea of an echo chamber (which suggests that opposing voices must be entirely removed and marginalised). 

Even though active climate change skeptic community members appear to have a clear stance on the issue, commenting and countering dissonant posts from outgroup members who accept the scientific consensus, we do not find evidence for more negative sentiment in these discussion threads. This is in line with qualitative findings that climate change skeptics appear to aim for an argument-based discourse culture in the forum, rather than one focused on ad-hominem attacks. An interesting and closely related characteristic of the discourse in r/climateskeptics, which we remarked upon during the coding procedure, is the emphasis on climate knowledge and overall competencies in STEM subjects among its members (e.g. `It's easy to claim a skeptic is a denier, if one doesn't even know the names of the most fundamental temperature sets.'). 

Our study also contributes to the literature on modern climate change skepticism \citep{lewandowsky_nasa_2013,lewandowsky2019science,edwards2013participants,van2020echo,walter_echo_2018,williams_network_2015,bruggemann2020mutual,elgesem2015structure,kaiser2017alliance} and mechanisms of online polarization \citep{sunstein2018republic,sunstein_law_2002,nordbrandt2020cross,bail_exposure_2018}. Most previous studies focus on the detection and description of polarized networks. In contrast, we examine concrete responses to opposing views, a theoretically implied but barely studied mechanism. While some studies view active engagement with opposing views as part of the solution to climate change skepticism, we find empirical evidence that it may in fact be part of the problem, at least in our context. 
Methodologically, while most previous studies either focused exclusively on the issue of climate change in online blogs (e.g. \citealt{matthews2015people}), established media outlets (e.g. \citealt{walter_echo_2018}) or the social media context (e.g. \citealt{williams_network_2015}), we use a methodological approach that draws upon insights from various online arenas. For example, the in-depth examination of submissions with manual content coding allowed us to understand how publications like the \textit{Guardian} can be used to support climate change skeptic positions, by presenting isolated facts without context and reframing them to support climate change skepticism or by using them in a sarcastic way to mock the `alarmism' of the outlet. 
One can only speculate whether climate change skeptics generally perceive themselves as minority opinion. However, findings on other minorities suggest that they experience higher epistemic needs to maintain their identity with intergroup discrimination than majority groups \citep{mullen1992ingroup,leonardelli2001minority}, which is in line with our findings of identity defence patterns in the community.

\subsection*{Limitations}

While we consider Reddit a good platform for our study, it is worth noting a key limitation of our focus here, which is that the sociodemographic characteristics of Reddit users are not comparable to the general population. During the registration process, no personal information is collected and the demographics of the 330 million active users remain largely unknown. However empirical investigations suggest that users are more likely to be male and younger than the general population \citep{duggan_6_2013,singer_evolution_2014}.

Another limitation was our reliance on manual coding. Nevertheless, given the highly context dependent and often ironic or sarcastic nature of submissions within r/climateskeptics, similar to the highly quantitative study of \citet{williams_network_2015}, a manual approach was chosen over automated content coding using NLP or machine learning techniques \citep{kirilenko_climate_2012,metag_content_2016,weitzel_comprehension_2016}. 

\subsection*{Conclusion}

A clear and possibly the most important finding of this study is, that in contrast to the classical theory of echo chambers, ‘breaking up the echo chamber’ with information on the consequences of climate change does not seem to work. While the major climate change skeptic community on Reddit, r/climateskeptics, is characterised by long-term stability, our study uncovered mechanisms of identity defence in reaction to opposing views and dissonant information. Therefore, we emphasise the need for the reconsideration of policy implications of promoting access to more cross cutting information. This line of reasoning proposes almost the opposite to the classical strategy of confronting problematic online communities with opposing information. However, further work is required to suggest more about what effective strategies are in such a context. 

Issues of conspiracy theory, involving lack of trust in scientists and politicians, have gained more importance when facing public reactions towards policies addressing the Covid-19 pandemic across various countries \citep{constantinou_is_2020}. When the consequences of climate change become equally pressing and visible, and more ambitious mitigation policies are implemented, the ground may be prepared for climate change skeptic conspiracy narratives to push into the mainstream with potentially catastrophic effects for effective policy making. It is therefore critical to understand and monitor climate change skepticism online in order to both gain insights about the discourse on climate change across the whole spectrum of the public sphere and to enable policy makers to act effectively when dynamics change.

\textbf{Lisa Oswald} is pursuing a PhD in Governance at the Hertie School in Berlin. She graduated from the University of Oxford, UK, with a MSc degree in Social Data Science, and from the University of Kassel, Germany, with a BSc and MSc degree in Psychology. She is interested in the public perception of climate change, political opinion formation, online communication dynamics and the emergence of collective behaviour. 

\textbf{Jonathan Bright} is an Associate Professor and Senior Research Fellow at the Oxford Internet Institute who specialises in computational approaches to the social and political sciences. He has two major research interests: exploring the ways in which new digital technologies are changing political participation; and investigating how new forms of data can enable local and national governments to make better decisions.

\textbf{Conflict of interest:} none.

\textbf{Supplementary Material:} available upon request. After publication available in data repository of respective journal.

\bibliography{references_sds_big}

\clearpage

\appendix

\section*{Appendix}

\section{Imputation} 
\label{app:imputation}

As described in the main body text we imputed values for submissions that could not be automatically labelled using our link based method. This occurred when the submissions did not contain a link or when the link was not contained in the coding dataset. In this case, the submission was given a value equal to the mean of all other submissions made by the author. In cases where the author themselves had no previously coded submissions, then these submissions were given a mean of the average submission value of all other authors. 

Of the 50,502 submissions, 8,645 (17\%) were imputed. 6,108 of these were imputed using the average of the author's submission history, and the remaining 2,537 were imputed using the average of all authors.  

All of the models reported in the text were also estimated on a subset of data without the imputed data points. The results were substantially the same: statistical significance and direction of effect was not altered, and size of effect was only altered fractionally. 

\section{Regression Diagnostics}
\label{app:diagnostics}

There are two types of models reported in the main text: our fixed effects regressions (M1.1 - M3.2, M5.1 and M5.2) and our PWP Gap Time model (M4.1). These two model types had two different diagnostic strategies. 

For the fixed effects regressions, we checked the distribution of errors for all models, and also whether there was any evidence of heteroscedasticity. This second check suggested that the use of robust standard errors would be advisable in some cases. In the end we have used these throughout for consistency (we make use of HC1 robust standard errors). We also checked for the presence of high leverage observations, of which at least some were found for most models. We used Cook's distance to perform this check: when points had a distance of greater than \( \frac{4}{Num. Obs}\) we labelled them as high leverage. We estimated further models without these high leverage observations, and the results were substantially the same.

For the Gap Time analysis, we also examined the potential impact of high leverage observations, using a cutoff of \( \frac{2}{\sqrt{Num. Obs}}\). Only one observation passed this threshold: a separate regression estimated without this observation did not show different results. Additionally, we estimated a further model which limited the amount of comments examined such that there were at least 50 per strata, a common correction for such models. Again, this model provided no major deviation from the main one reported in the text. Finally, we checked the proportional hazards assumption of the model. The results of this check are described in the main text in figure \ref{prop_haz}.

For all models, we also estimated versions where text only submissions were excluded (i.e. those without a link) and where submissions whose dissonance code was imputed were excluded. Again, this did not make a substantive difference to the results. 

\end{document}